\documentclass[prl,twocolumn,superscriptaddress,showpacs]{revtex4-1}
\usepackage{graphicx}
\usepackage{hyperref}
\usepackage{multirow}
\usepackage{fontenc}
\usepackage{amsmath}

\begin{document}

\title{Low-Energy Polymeric Phases of Alanates}
\author{Tran Doan Huan}
\affiliation{Department of Physics, Universit\"{a}t Basel, Klingelbergstrasse 82, 4056 Basel, Switzerland}
\author{Maximilian Amsler}
\affiliation{Department of Physics, Universit\"{a}t Basel, Klingelbergstrasse 82, 4056 Basel, Switzerland}
\author{Miguel A. L. Marques}
\affiliation{Universit\'e de Lyon, F-69000 Lyon, France and LPMCN, CNRS, UMR 5586, Universit\'e Lyon 1, F-69622 Villeurbanne, France}
\author{Silvana Botti}
\affiliation{Universit\'e de Lyon, F-69000 Lyon, France and LPMCN, CNRS, UMR 5586, Universit\'e Lyon 1, F-69622 Villeurbanne, France}
\author{Alexander Willand}
\affiliation{Department of Physics, Universit\"{a}t Basel, Klingelbergstrasse 82, 4056 Basel, Switzerland}
\author{Stefan Goedecker}
\email{stefan.goedecker@unibas.ch}
\affiliation{Department of Physics, Universit\"{a}t Basel, Klingelbergstrasse 82, 4056 Basel, Switzerland}

\date{\today}

\begin{abstract}
Low-energy structures of alanates are currently known to be described by patterns of isolated, nearly ideal tetrahedral [AlH$_4$] anions and metal cations. We discover that the novel polymeric motif recently proposed for LiAlH$_4$ plays a dominant role in a series of alanates, including LiAlH$_4$, NaAlH$_4$, KAlH$_4$, Mg(AlH$_4$)$_2$, Ca(AlH$_4$)$_2$ and Sr(AlH$_4$)$_2$. In particular, most of the low-energy structures discovered for the whole series are characterized by networks of corner-sharing [AlH$_6$] octahedra, forming wires and/or planes throughout the materials. Finally, for Mg(AlH$_4$)$_2$ and Sr(AlH$_4$)$_2$, we identify two polymeric phases to be lowest in energy at low temperatures.

\end{abstract}


\maketitle

Hydrogen is a compelling alternative to fossile fuels as it can deliver clean energy and is readily available in large quantities. Solid state hydrides for hydrogen storage as intended on board of vehicles need to provide high gravimetric hydrogen density ($\gtrsim 5.5$ \% wt), reasonable decomposition temperatures ($ \lesssim 100 ^\circ$ C), and full reversibility~\cite{schlapbach,mandal}. 
Tremendous research efforts have been devoted to metal alanates M(AlH$_4$)$_n$ (where M is a metal cation of valence $n$) as promising candidates. At low temperatures, alanates are known to crystallize in phases which are characterized by patterns of isolated, nearly ideal tetrahedral [AlH$_4]^-$ anions and M$^{n+}$ cations \cite{hauback02Li, lovvik04Li, bogdanovic97Na, zaluska00Na, opalka03Na, hauback03Na, wood, morika03K, hauback05K, vajeeston04K, wol07Ca, fichtnerMg, lovvik05Mg, Palumbo07Mg, pommerin12Sr}. 
Recently, low-energy phases were  predicted for LiAlH$_4$ with a novel structural motif \cite{amsler12}. In these phases, networks of corner-sharing [AlH$_6$] octahedra form wires and planes throughout the material. We will hereafter refer to such  phases as ``polymeric phases" while the expression ``isolated phases" will be used for structures characterized by isolated [AlH$_4$] tetrahedra.

The discovery of energetically favorable polymeric phases in LiAlH$_4$ may have an impact on the controversy over the stability of this compound \cite{lovvik04Li, KangLi04, Dymova}. Furthermore, the [AlH$_6$] octahedra may influence the kinetics of the dehydrogenation process~\cite{Dymova279, lovvikDecom}
\begin{equation}\label{decomp1}
3{\rm MAlH_4} \to {\rm M_3AlH_6 + 2Al + 3H_2},\; {\rm (M=Li, Na, K),}
\end{equation}
since the same [AlH$_6$] octahedra have been reported in Li$_3$AlH$_6$~\cite{lovvik04Li, VajesstonLi04}. This behavior could be transferred to other alanates and may be helpful in for example improving their stability at ambient conditions, e.g. Mg(AlH$_4$)$_2$ \cite{lovvik05Mg, Palumbo07Mg}. Also, the [AlH$_6$] octahedra have been observed in the dehydrogenation products of many other alanates, such as Na$_3$AlH$_6$ \cite{ronnebro00Na} and K$_3$AlH$_6$ \cite{morika03K, lovvikDecom} from NaAlH$_4$ and KAlH$_4$ via reaction~\eqref{decomp1}, and CaAlH$_5$ and SrAlH$_5$ from Ca(AlH$_4$)$_2$ and Sr(AlH$_4$)$_2$ via reaction~\eqref{decomp2}, respectively~\cite{fichnerCa, wol07Ca, MamathaCa, klaveness,Dymova531, pommerin12Sr}
\begin{equation}\label{decomp2}
2{\rm M(AlH_4)_2} \to {\rm 2MAlH_5+2Al+3H_2}, {\rm (M=Ca, Sr).}
\end{equation}
These similarities strongly suggest that polymerization of [AlH$_4$] in alanates may influence the dehydrogenation processes of such compounds.

In this Letter, we investigate the low-energy polymeric phases of a series of six alanates, including three alkali metal alanates LiAlH$_4$, NaAlH$_4$, KAlH$_4$, and three alkaline earth metal alanates Mg(AlH$_4$)$_2$, Ca(AlH$_4$)$_2$, Sr(AlH$_4$)$_2$. The first-principles calculations in this work were performed at the density functional theory (DFT) \cite{dft1, dft2} level using the {\sc abinit} package \cite{abinit1, abinit2,GonzePhonon}. We used the generalized gradient approximation (GGA), the Perdew-Burke-Ernzerhof (PBE) exchange-correlation functional \cite{per96} and the norm-conserving Hartwigsen-Goedecker-Hutter pseudopotentials \cite{HGH} for total energy and linear response phonon calculations. The plane-wave cutoff energy was 60 hartree (Ha) while a Monkhorst-Pack ${\bf k}$-point mesh \cite{monkhorst} was chosen for each structure, ensuring the convergence of the total energy to be better than $10^{-5}$ Ha/atom. Atomic and cell variables were simultaneously relaxed until all the residual force and stress components were smaller than $10^{-5}$ Ha $\times$ bohr$^{-1}$ and $10^{-7}$ Ha $\times$ bohr$^{-3}$, respectively. Additional calculations with PBEsol \cite{PBEsol} and LDA exchange-correlation functionals were performed to confirm the energetic orderings of the phases.

\begin{figure}[t]
  \begin{center}
    \includegraphics[width=8.25cm]{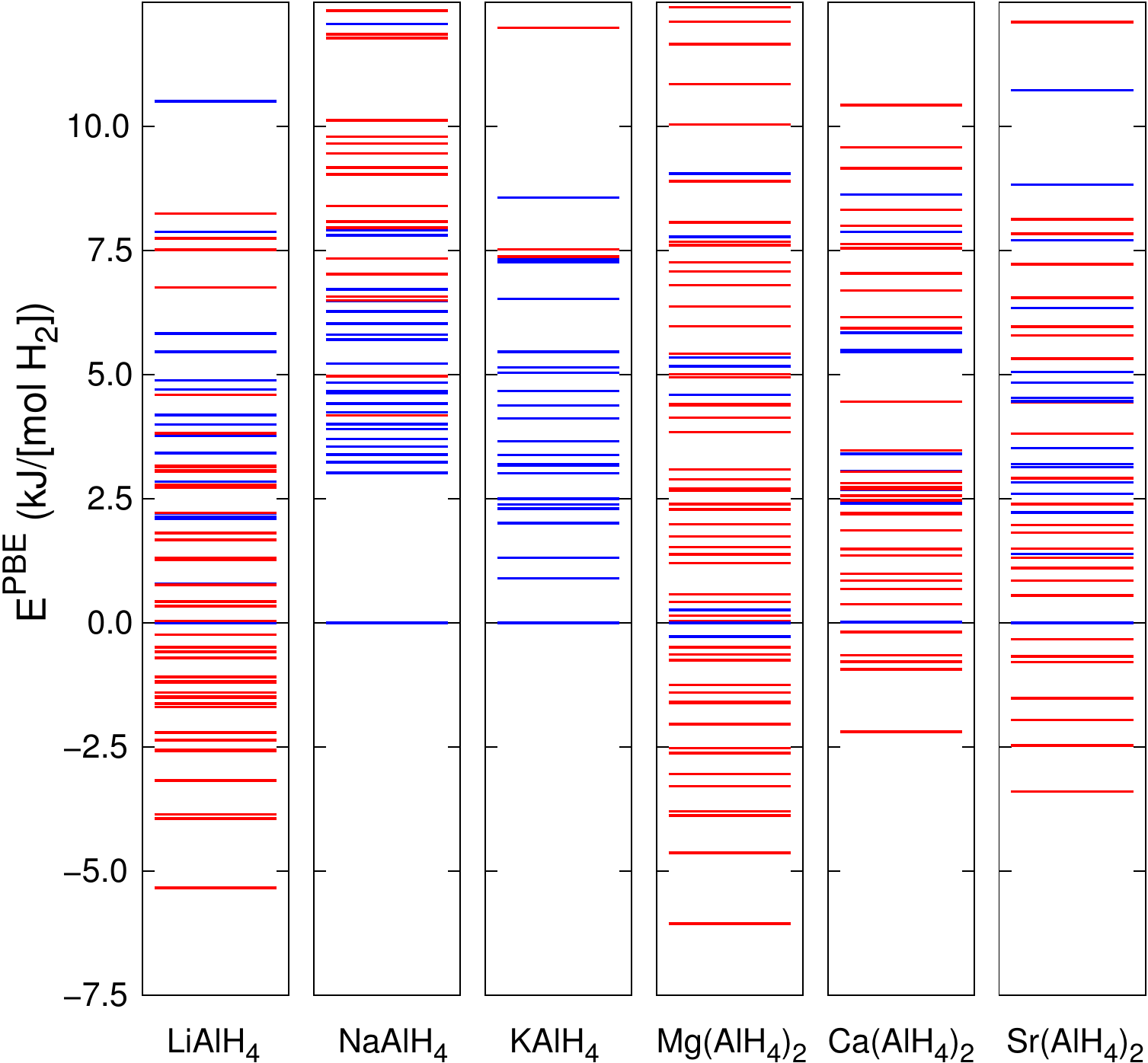}
  \caption{(Color online) Low-energy spectra of LiAlH$_4$, NaAlH$_4$, KAlH$_4$, Mg(AlH$_4$)$_2$, Ca(AlH$_4$)$_2$ and Sr(AlH$_4$)$_2$. Blue/red lines represent the isolated/polymeric phases. Energies are given in unit of kJ/[mol H$_2$] with respect to those of the reference phases (see the text).}\label{fig:spectrum}
  \end{center}
\end{figure}

The recently generalized minima-hopping method (MHM)~\cite{goedecker04,amsler10} was used to predict novel crystal structures. The DFT-energy landscape is thereby explored by short consecutive molecular dynamics steps followed by local geometry relaxations. The initial velocities for the molecular dynamics runs are chosen approximately along soft mode directions, allowing efficient escapes from local minima, and aiming toward the global minimum. The predictive power of the MHM has already been demonstrated in a wide range of applications~\cite{hellmann_2007, roy_2009, bao_2009, willand_2010, de11, amsler_crystal_2011, Livas, amsler12}.

\begin{table}[t]
\begin{center}
\caption{Summary of the three most stable polymeric phases of each alanate. Total energies $E^{\rm PBE}$, $E^{\rm PBEsol}$ and $E^{\rm LDA}$ are obtained with PBE, PBEsol, and LDA exchange-correlation functionals. Zero-point energies $E_{\rm ZP}$ are included in $E^{\rm PBE}_{\rm ZP}$. The energies are given in unit of kJ/[mol H$_2$] with respect to the reference phases. Space group numbers are given in parentheses.}\label{table_energy}
\begin{tabular}{llrrrr}
Compound & Space group &$E^{\rm PBE}$ & $E^{\rm PBE}_{\rm ZP}$& $E^{\rm PBEsol}$& $E^{\rm LDA}$\\
\hline
LiAlH$_4$    & $P2_1/c$ (14)     & $-5.32$ & $-3.07$ & $-12.15$ & $-12.38$ \\
              & $P2_1$ (4)        & $-3.94$ & $-1.65$ & $-11.12$ & $-11.33$ \\
              & $Pnc2$ (30)       & $-3.92$ & $-1.49$ & $-11.44$ & $-11.72$ \\
 \hline
NaAlH$_4$      &  $C2/m$ (12)         & $4.07$ & $5.28$ &$3.31$  & $2.67$ \\
                &  $P\overline{1}$ (2) & $5.03$ & $6.66$ & $3.89$ & $3.81$\\
                &  $C2$ (5)            & $6.30$ & $7.57$ & $5.32$ & $0.21$ \\
 \hline
KAlH$_4$    &  $P\overline{1}$ (2) & $7.37$  & $10.45$ & $0.84$ & $-0.37$\\
             &  $Ama2$ (40)         & $7.56$  & $10.00$ & $3.70$ & $3.60$\\
             &  $Cmcm$ (63)         & $11.99$ & $14.01$ & $5.56$ & $3.69$\\
 \hline
Mg(AlH$_4$)$_2$ &  $P2_1$ (4)  &$-6.07$ & $-2.53$ & $-16.29$  & $-18.38$\\
                 &  $P2$ (3)    &$-4.59$ & $-0.79$ & $-13.88$ & $-16.74$\\
                 & $C2/m$ (12)  &$-3.89$ & $-0.09$ & $-12.84$ & $-15.07$\\
 \hline
Ca(AlH$_4$)$_2$ &  $P2_1/c$ (14) & $-2.17$ & $0.48$ &$-12.15$ & $-14.26$\\
                 &  $C2$ (5)      & $-0.85$ & $1.72$ &$-11.04$ & $-13.19$\\
                 &  $Pm$ (6)      & $-0.77$ & $1.86$    &$-11.74$ & $-14.54$\\
 \hline
Sr(AlH$_4$)$_2$ & $Pm$ (6)             & $-3.49$ & $-1.18$     & $-8.94$ & $-9.70$ \\
                 & $P2_1/c$ (14)        & $-2.56$ & $-0.67$ & $-5.85$ & $-5.70$ \\
                 & $P\overline{1}$ (2)  & $-2.05$ & $0.38$  & $-6.75$ & $-7.30$ \\
\hline
\end{tabular}
\end{center}
\end{table}

\begin{figure*}[t]
  \begin{center}
    \includegraphics[width=13 cm]{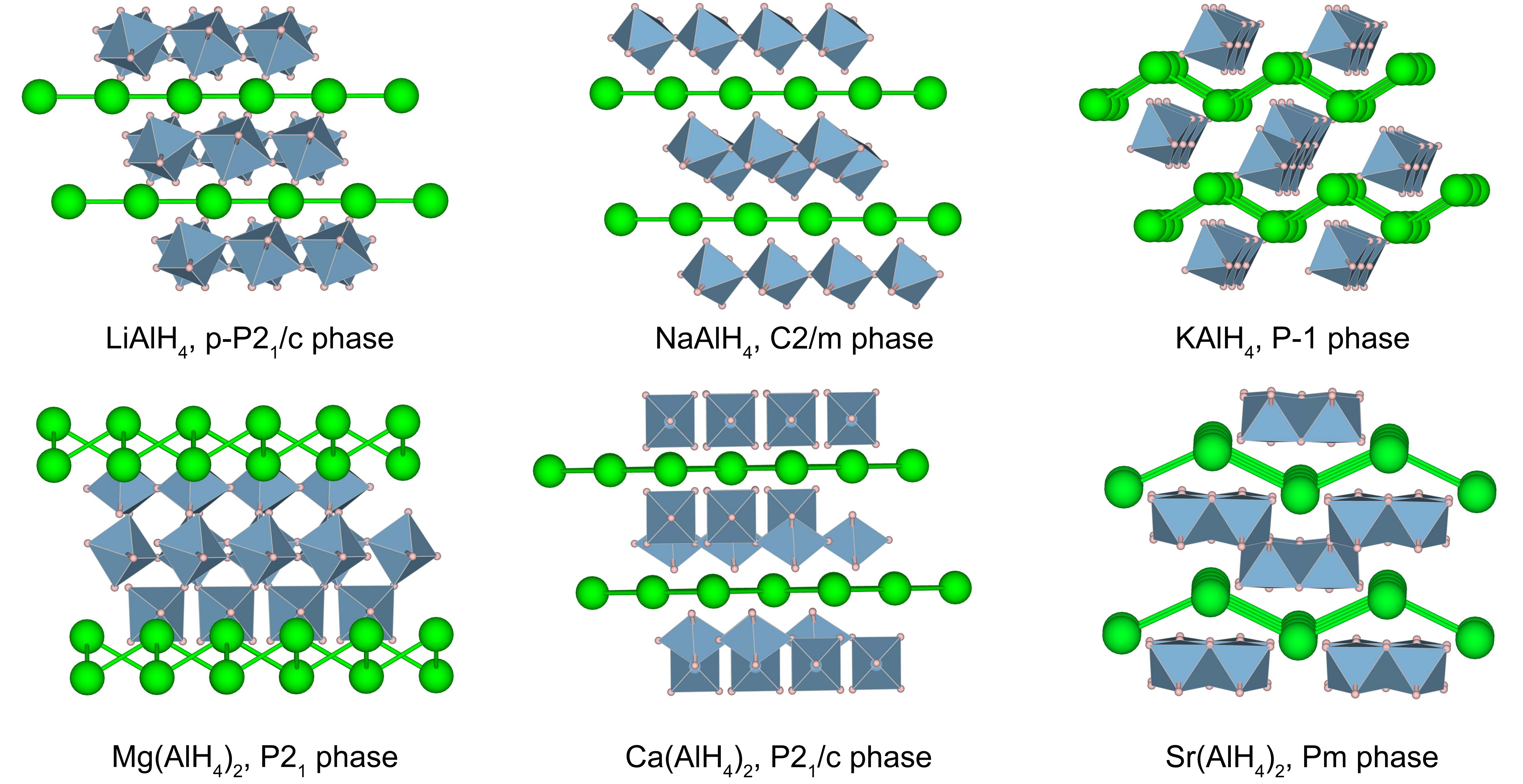}
  \caption{(Color online) Most stable polymeric phases of LiAlH$_4$, NaAlH$_4$, KAlH$_4$, Mg(AlH$_4$)$_2$, Ca(AlH$_4$)$_2$, and Sr(AlH$_4$)$_2$. Green spheres represent metal cations while octahedra are formed by [AlH$_6$] complexes. Space groups are also given.}\label{fig:struct}
  \end{center}
\end{figure*}

We performed several MHM simulations with up to 4 formula units (f.u.) per cell for LiAlH$_4$, NaAlH$_4$ and KAlH$_4$, and with up to 2 f.u. per cell for Mg(AlH$_4$)$_2$, Ca(AlH$_4$)$_2$, and  Sr(AlH$_4$)$_2$, starting from random input structures. For a given number of formula units per primitive cell, our runs were able to recover all of the previously reported isolated phases of the alanates, which belong to the space group $P2_1/c$ for LiAlH$_4$ \cite{hauback02Li,lovvik04Li}, $I4_1/a$ \cite{hauback03Na} and $Cmcm$ \cite{wood} for NaAlH$_4$ , $Pnma$ for KAlH$_4$ \cite{hauback05K}, and $P\overline{3}m1$ for Mg(AlH$_4$)$_2$ \cite{fichtnerMg, lovvik05Mg}. 
The $Pcba$ phase for Ca(AlH$_4$)$_2$ \cite{wol07Ca} could obviously not be found since it has 8 f.u. per primitive cell. For Sr(AlH$_4$)$_2$, no conclusive crystal structure is available in literature to our best knowledge~\cite{ pommerin12Sr}. 

Furthermore, we discovered a large number of novel low-energy structures for all the compounds. The energy spectra of these phases are shown in Fig.~\ref{fig:spectrum} (without zero-point correction $E_{\rm ZP}$) with respect to the reference structures, chosen to be the most stable reported isolated phases. For Sr(AlH$_4$)$_2$, the lowest isolated phase (space group $P\overline{1}$) from our predictions in Sr(AlH$_4$)$_2$ was used as reference~\cite{supplement}. Although several polymeric phases were found for NaAlH$_4$ and KAlH$_4$, their most stable phases are isolated. On the other hand, a large number of polymeric phases were discovered for LiAlH$_4$, Mg(AlH$_4$)$_2$, Ca(AlH$_4$)$_2$, and Sr(AlH$_4$)$_2$, dominating their low-energy configurational space. The geometry of the most stable polymeric phase for each alanate is shown in Fig.~\ref{fig:struct}.

\begin{figure}[b]
  \begin{center}
    \includegraphics[width=8.25cm]{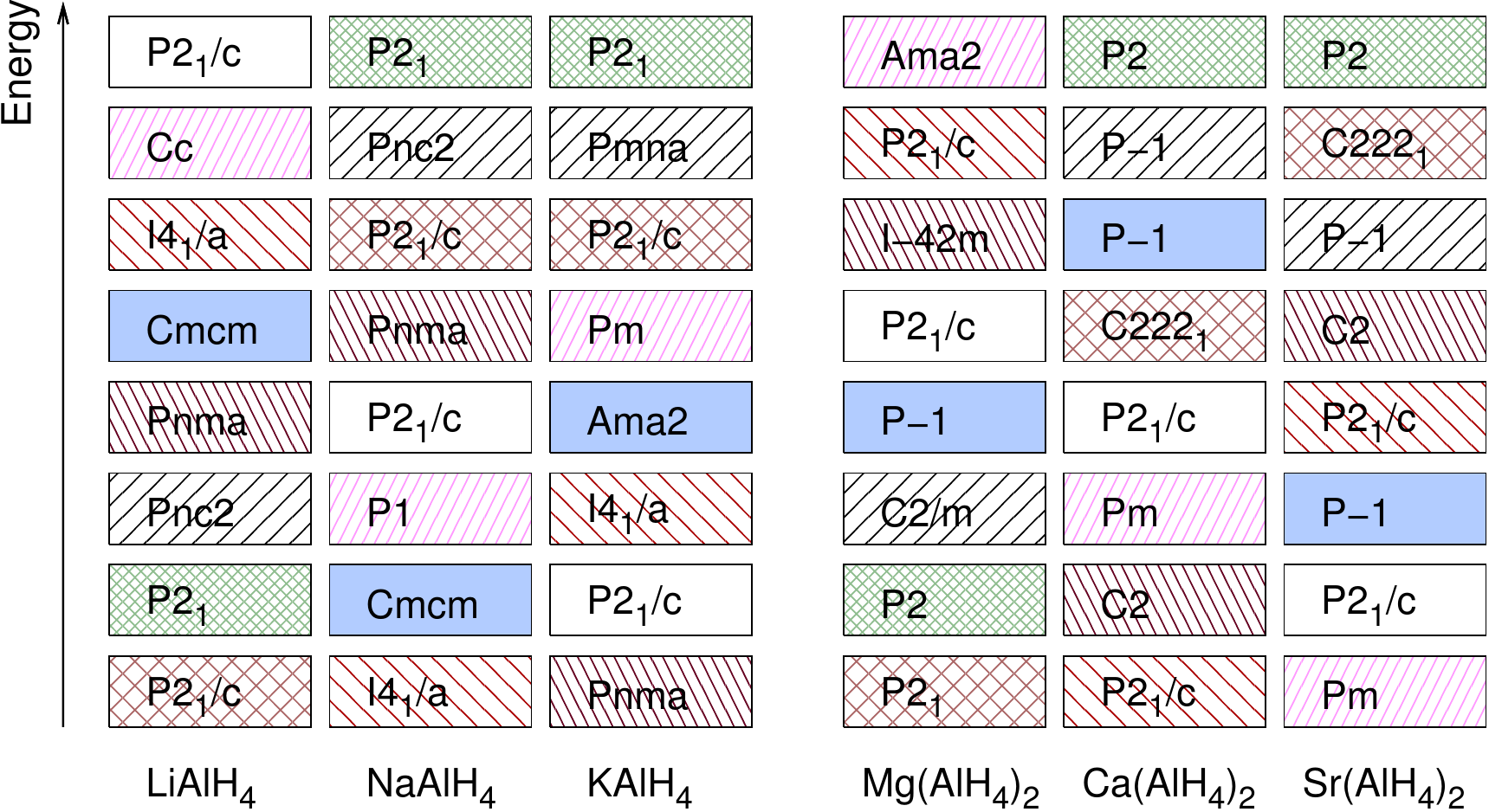}
  \caption{(Color online) Energetic ordering and space groups of the structures obtained by substituting the cations of the alanates within a group. For each of the two groups (M=Li, Na, K and M=Mg, Ca, Sr) structures with identical colors and patterns have the same origin.}\label{fig:diagram}
  \end{center}
\end{figure}

A summary of the energetic and structural properties of the most stable polymeric phases is given in Table \ref{table_energy} (see also Supplemental Material \cite{supplement}). For these structures, the total energies $E^{\rm PBE}$, $E^{\rm PBEsol}$, and $E^{\rm LDA}$ were obtained with the PBE, PBEsol, and LDA exchange-correlation functionals, respectively, while $E_{\rm ZP}$ was calculated with the PBE functional and included in $E^{\rm PBE}_{\rm ZP}$. Without $E_{\rm ZP}$, the energetic orderings are invariant with respect to the different employed exchange-correlation functionals. However, $E_{\rm ZP}$ was found to be important and may change the energetic orderings, i.e., with $E_{\rm ZP}$, the most stable phases of NaAlH$_4$ ($I4_1/a$), KAlH$_4$ ($Pnma$), and Ca(AlH$_4$)$_2$ ($Pbca$) are isolated while the most stable phases of LiAlH$_4$ ($P2_1/c$), Mg(AlH$_4$)$_2$ ($P2_1$), and Sr(AlH$_4$)$_2$ ($Pm$) are polymeric. Although GGA  systematically underestimate the decomposition enthalpy of many compounds by $\sim 10$ kJ/mol H$_2$ \cite{akbarzadeh, peles}, it is expected to play no role in the energetic stability of isolated versus polymeric phases.  The simulated X-ray diffraction spectra were compared to experimental results and can be found in the Supplemental Materials \cite{supplement}.

In polymeric phases, the negatively charged [AlH$_6$] octahedra are linked together by sharing their corners to form networks of wires and/or planes throughout the material. The octahedra in alanates share the same geometrical structure: the Al atom at the center is surrounded by four H atoms on a plane while the other two H atoms form Al-H bonds nearly perpendicular to the plane. To estimate the charge transfer onto the [AlH$_6$] polymers, we performed a Bader charge analysis calculation on the $P2_1$ phase of Mg(AlH$_4$)$_2$ and on the $Pm$ phase of Sr(AlH$_4$)$_2$ using the {\sc aim} utility of the {\sc abinit} package. We found that a charge of $-1.65q_e$ and $-1.36q_e$ is transferred from a Mg and Sr atom to the [AlH$_6$] complexes, respectively, clearly indicating that the polymeric sub-structures themselves carry a strong electric charged.

The [AlH$_6$] octahedral geometry, shown \cite{peles} to be a consequence of the hybridization of the H $1s$ and Al $3d$ orbitals in Na$_3$AlH$_6$, is a common feature of the alanates. This geometry clearly distinguishes the alanates from complex borohydrides, e.g., LiBH$_4$, of which only [BH$_4$] tetrahedra were observed. In such compounds, the similar [BH$_6$] octahedral geometry is unfavorable because the $3d$ state has a much higher energy than the $2s$ and $2p$ states in B \cite{peles}.

\begin{figure}[b]
  \begin{center}
    \includegraphics[width=8.25cm]{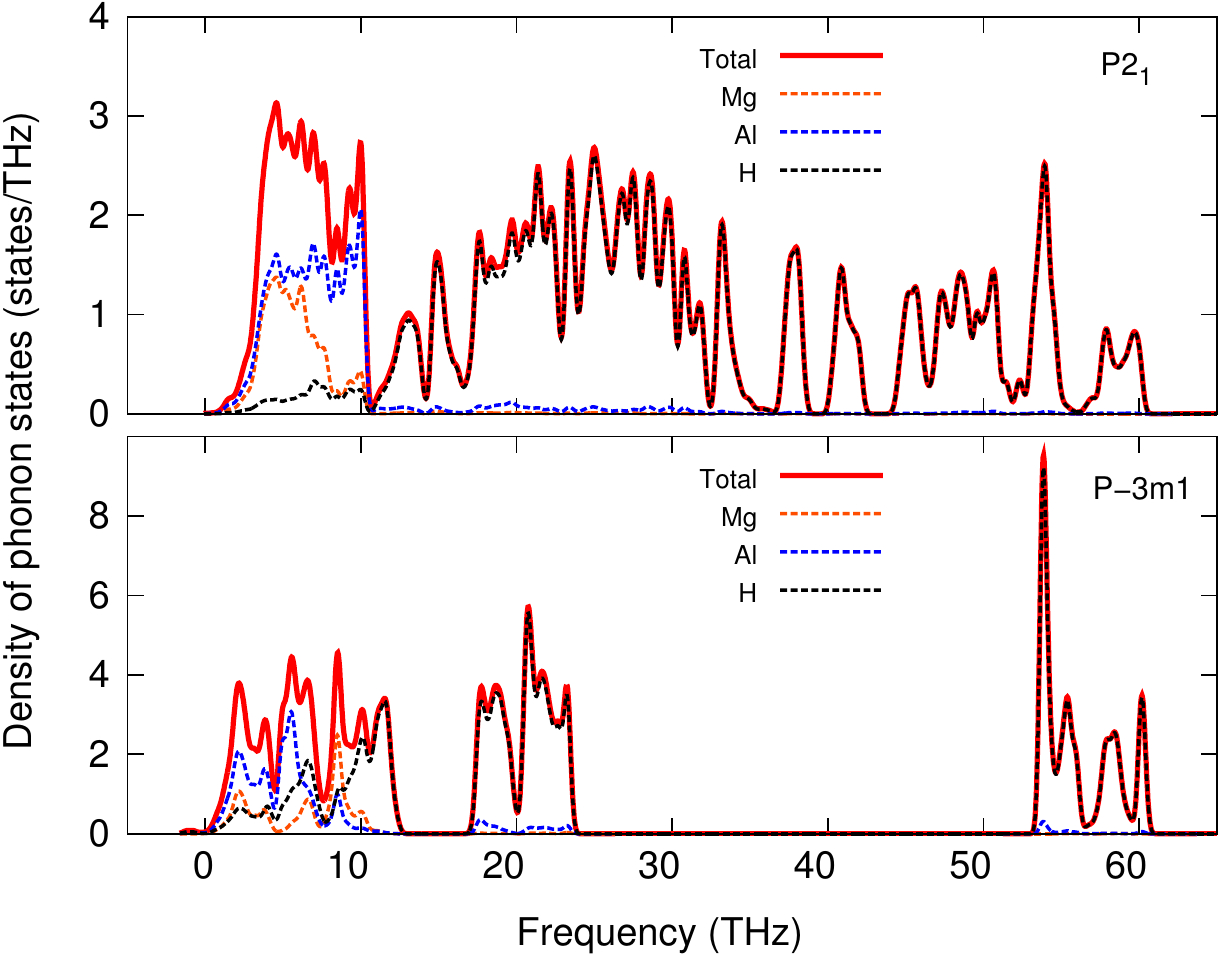}
  \caption{(Color online) Densities of phonon states (partial and total) for 2 f.u. of Mg(AlH$_4$)$_2$ in $P2_1$ and $P\overline{3}m1$ phases.}\label{fig:phdos}
  \end{center}
\end{figure}

We further analyzed structural and energetic relationships between the alanates. For three lowest-energy phases of each alanate, we replaced the cation by the two other cations of same valence, and then a small step-size local geometry relaxation was performed. The energetic ordering of the relaxed structures is shown in Fig. \ref{fig:diagram}. There is no clear rank correlation of the three alanates within a group. We obtained a geometric average of $\tau = 0.47$ for the Kendall tau rank correlation for the group of monovalent metal alanates and  $\tau = 0.44$ for the divalent group. In particular, the ground state structures are completely unrelated with each other. Therefore, database searching methods which are often based on exchanging cations of ground state phases in another compound are bound to fail to predict the correct structure, emphasizing the need for unconstrained and systematic structure prediction algorithms. 

In order to investigate the dynamical stability we performed linear response phonon calculations for all the polymeric phases shown in Table \ref{table_energy}. No imaginary phonon modes were observed in the whole Brillouin zones, indicating that all phases are dynamically stable. The zero-temperature densities of phonon states $\rho_{\rm ph}(\omega)$ of these phases are given in the Supplemental Material \cite{supplement}. The partial and total densities of phonon states of the polymeric $P2_1$ phase and the isolated $P\overline{3}m1$ phase of Mg(AlH$_4$)$_2$ are compared in Fig.~\ref{fig:phdos}. In both phases, three frequency ranges (below $12$ THz, $12-40$ THz, and above $40$ THz) correspond to the Mg/Al framework vibrations, the molecular libration and Al-H bending modes, and the Al-H stretching modes, respectively. Because the [AlH$_4$] tetrahedra in the $P\overline{3}m1$ phase are isolated and essentially independent, these frequency ranges are localized in narrow energy windows and are clearly distinct. On the other hand, the librational/Al-H bending frequencies and the Al-H stretching frequencies of the $P2_1$ phase are smeared out in a wide energy range. This behavior is a consequence of the polymeric motifs of the [AlH$_6$] octahedra, which are linked together by the H atoms at several corners. Therefore, the contribution of the zero-point vibrational energy at 0~K of any polymeric phase, quantified by  $E_{\rm ZP} \equiv \int_0^\infty d\omega\left[\frac{\omega}{2} \rho_{\rm ph}(\omega) \right]$, is larger than in isolated phases, as shown in Table \ref{table_energy}.

\begin{figure}[t]
  \begin{center}
    \includegraphics[width=8.25cm]{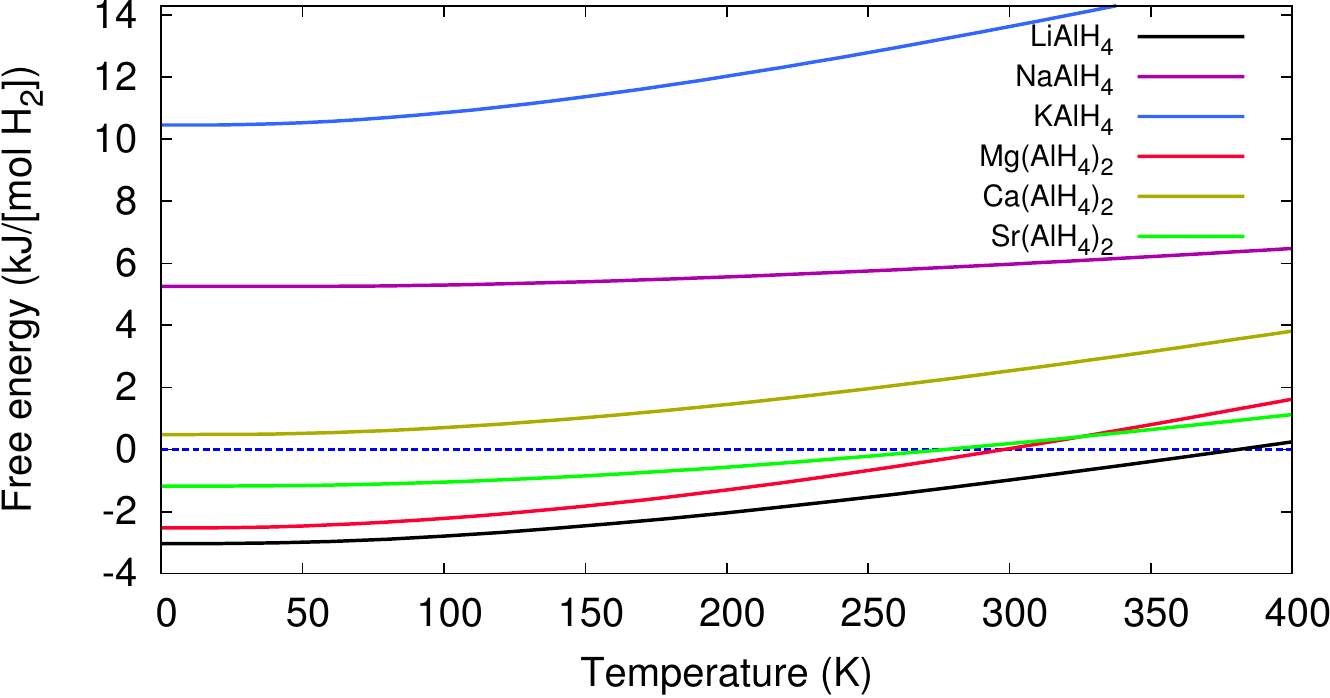}
  \caption{(Color online) Free energy of the lowest polymeric structures of the alanates, given with respect to that of the corresponding isolated reference structures.}\label{fig:free_energy}
  \end{center}
\end{figure}

At finite temperatures, the significance of the vibrational energy becomes even more apparent. Fig. \ref{fig:free_energy} shows the free energies of the most stable polymeric phases of the alanates, given with respect to the reference phases. All polymeric phases become monotonically less (thermodynamically) stable as the temperature increases. This behavior demonstrates that the vibrational contribution to the free energy of any polymeric phase grows faster than in the corresponding isolated reference  phase. For LiAlH$_4$, Mg(AlH$_4$)$_2$, and Sr(AlH$_4$)$_2$, the $P2_1/c$, $P2_1$, and $Pm$ phases become less stable than the corresponding reference phases above $380$K, $300$K, and $280$ K, respectively. On the other hand, the $P2_1$ phase of Mg(AlH$_4$)$_2$ could enhance the stability of Mg(AlH$_4$)$_2$ which has been reported to be metastable at ambient conditions in the $P\overline{3}m1$  phase~\cite{lovvik05Mg, Palumbo07Mg}.

\begin{figure}[t]
  \begin{center}
    \includegraphics[width=8.25 cm]{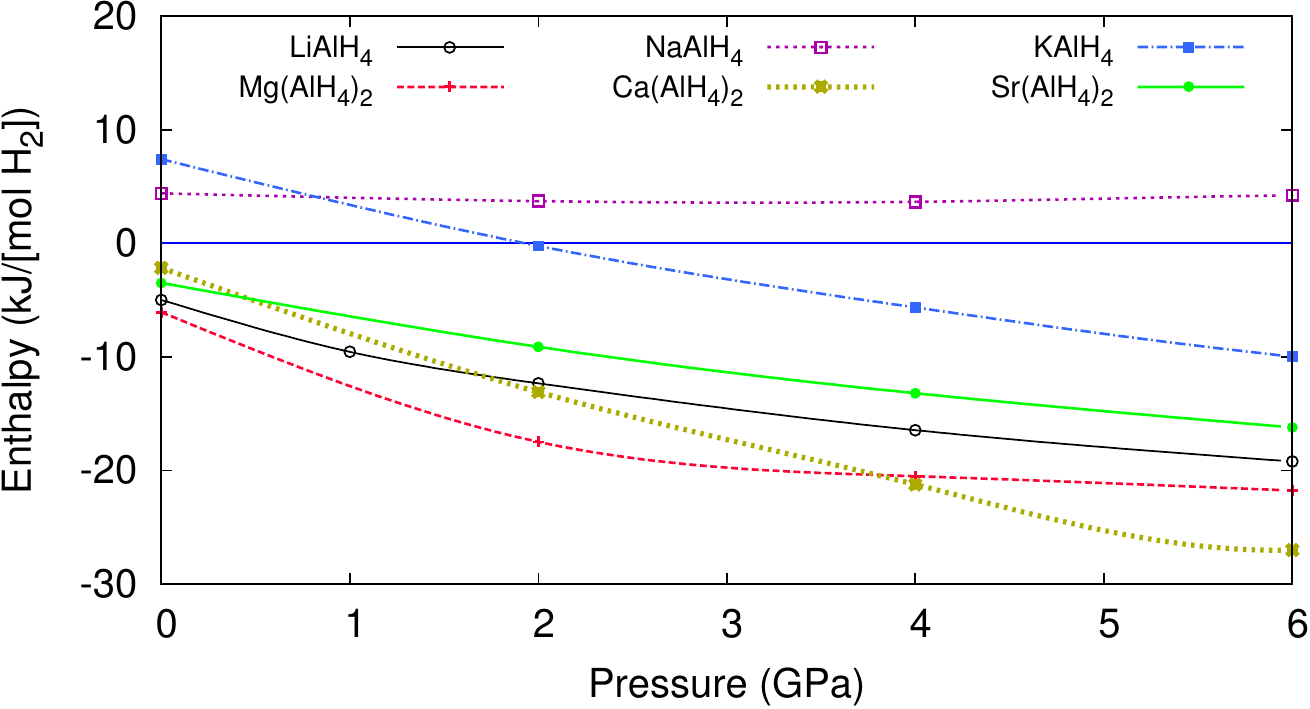}
  \caption{(Color online) Enthalpies as a function of pressure for the lowest polymeric structures with respect to the corresponding isolated reference phases.}\label{fig:enthalpy}
  \end{center}
\end{figure}

Furthermore, we investigated the electronic properties of two polymeric phases and performed $GW$ calculations~\cite{GWA} for the $P2_1$ phase of Mg(AlH$_4$)$_2$ and the $Pm$ phase of Sr(AlH$_4$)$_2$, and their corresponding reference phases. The polymeric $P2_1$ phase was found to be an insulator with an indirect band-gap of 4.7 eV, which is significantly lower than that of the isolated $P\overline{3}m1$ phase, determined to be 6.5 eV in agreement with previous results \cite{lovvik05Mg,Palumbo07Mg}. For Sr(AlH$_4$)$_2$, we found that the $Pm$ phase is also an insulator with a gap of 5.3 eV.	

Fig.~\ref{fig:enthalpy} demonstrates that the profound differences in the structural motifs of the polymeric and isolated phases also have a strong influence on their compressibility and relative thermodynamical stability upon compression. It shows the pressure evolution of the enthalpies of the lowest polymeric structures with respect to the reference phases. As already indicated in Ref.~\cite{amsler12} for the case of LiAlH$_4$, the stability of the polymeric phases improves drastically as the pressure increases, with the exception of NaAlH$_4$. 
In particular, the isolated phase in KAlH$_4$ is predicted to transform into a polymeric phase at a pressure of 2~GPa and 0~K. Therefore, cold compression of alanates could be a promising approach en route to experimental synthesis of polymeric phases. Furthermore, the polymeric phases may be helpful in bringing some alanates with small decomposition enthalpies, e.g., LiAlH$_4$ with $\sim 9$ kJ/mol H$_2$ \cite{lovvik04Li, akbarzadeh}, to approach the target range (30-45 kJ/mol H$_2$), especially under pressure.

In conclusion, we unveiled the complexity of the energy landscape in several alanates and the dominant role of the recently-proposed polymeric phases in their low-energy polymorphs. These polymeric phases, a common feature of the alanates, are characterized by a charged network of corner-sharing [AlH$_6$] octahedra and a significant vibrational energy. Two insulating polymeric phases of $P2_1$ and $Pm$ symmetries were identified to be the most stable in Mg(AlH$_4$)$_2$ and Sr(AlH$_4$)$_2$. Both free energy and enthalpy calculations show that such phases should in fact be possible to synthesize, either at low temperatures or at high pressures. Further experimental investigations are necessary for a deeper understanding of such structural properties in alanates.

\begin{acknowledgments}
The authors thank Nguyen-Manh Duc and Nicola Marzari for useful discussions. TDH, MA, AW, and SG acknowledge the financial support provided by the Swiss National Science Foundation. Computational resources were provided by the Swiss National Supercomputing Center in Lugano.
\end{acknowledgments}

\end{document}